\documentclass{elsart5p}
\usepackage{exscale}
\usepackage{graphics}
\usepackage{graphicx}
\usepackage{amssymb,bm}
\def\sgn{\mbox{sgn}}
\begin{document}

\begin{frontmatter}

\title{Critical Kondo destruction and the violation of the 
quantum-to-classical mapping of quantum criticality}

\author[AA]{Stefan Kirchner\corauthref{Name1}},
\ead{kirchner@rice.edu}
\author[AA]{Qimiao Si},
\address[AA]{Department of Physics \& Astronomy, Rice University, Houston,
TX 77005, USA}
\corauth[Name1]{Corresponding author. Tel: (713) 348-4291 fax: (713)348-4150}

\begin{abstract}
Antiferromagnetic heavy fermion metals close to their quantum critical points display a richness 
in their physical properties unanticipated by the traditional approach
to quantum criticality, which
describes the critical properties solely in terms of fluctuations of the order parameter. 
This has led to the question as to how the Kondo effect gets destroyed as the system undergoes 
a phase change. In one approach to the problem, Kondo lattice systems are studied through a 
self-consistent Bose-Fermi Kondo model within the Extended Dynamical Mean Field Theory. The 
quantum phase transition of the Kondo lattice is thus mapped onto that of
a  
sub-Ohmic  Bose-Fermi Kondo model.
In the present article we address some aspects of the failure of 
the standard order-parameter functional for the
the Kondo-destroying quantum critical point of the 
Bose-Fermi Kondo model.
\end{abstract}


\end{frontmatter}

\section{Introduction}
Heavy fermion systems have attracted considerable interest ever since 
the initial studies \cite{Stewart.84} in the 1970s.
While the early interest was fueled by the unusual 
Fermi liquid properties and superconductivity
in these 
naturally magnetic materials,
recent focus 
has been 
on their
quantum critical behavior.
In the following, we will focus on 
antiferromagnetic heavy fermion metals, 
in which a zero-temperature phase transition between 
a paramagnet and an itinerant antiferromagnet
occurs when an external tuning parameter, 
{\it e.g.} external pressure, magnetic field,
or compound composition, is varied~\cite{Gegenwart.08,Loehneysen.07}.
This is made possible in part by the fact that the magnetic
energy scales in these systems are small enough 
so that a moderate change of the tuning parameter 
corresponds to a sizeable perturbation.
Because they are metallic,
these materials
allow for systematically 
addressing the interplay between quantum critical
fluctuations and unusual electronic excitations.

The traditional approach to an itinerant antiferromagnetic 
quantum critical point (QCP)
describes its universal properties in terms of a 
Ginzburg-Landau-Wilson functional of the order parameter 
and its fluctuations 
-- the $\phi^4$ theory -- in $d+z$ dimensions~\cite{Hertz.76,Millis.93}, 
where $d$ is the spatial dimension (typically $3$ or $2$)
and $z$ corresponds to the dynamic exponent.
Because the effective dimensionality is at or above 
the upper critical dimension of the $\phi^4$ theory,
one is led to expect Gaussian behavior for such 
a Hertz-Millis or T=0 spin-density wave 
(SDW) QCP. 
The anticipated behavior of the spin relaxation rate $\Gamma_{Q}$
at the ordering wave vector is (for $d=3$) $\Gamma_{Q}\sim T^{3/2}$,
which has indeed been observed in one heavy fermion
system~\cite{Kadowaki.06}. There are however a number
of observations that cannot be explained within the SDW
approach~\cite{Schroeder.00,Paschen.04,Park.06,Gegenwart.07}.  
In $CeCu_{6-x}Au_{x}$, at its critical concentration $x_c \approx 0.1$,
the spin relaxation rate is linear in temperature, 
and the order-parameter susceptibility 
singularly depends on
temperature and frequency 
with a fractional exponent~\cite{Schroeder.00}.
In $YbRh_2Si_2$, multiple energy scales collapse 
at its field-tuned QCP and, moreover, 
the Fermi surface has been implicated to 
transition from large to small when the QCP is 
crossed from the paramagnetic side.
These rich behaviors have led to the question as to 
whether and how the Kondo screening is affected
as the system undergoes a magnetic phase 
change~\cite{Si.01,Coleman.01}.

\begin{figure}[t!]
\begin{center}
\includegraphics[width=0.5\textwidth]{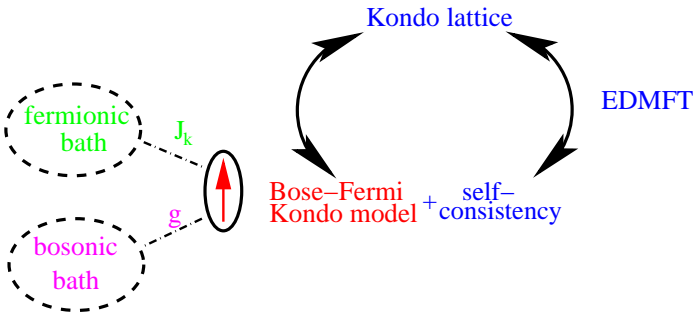}
\end{center}
\caption{EDMFT scheme: The Kondo lattice model is mapped to
a quantum impurity model, 
the Bose-Fermi Kondo model, augmented by a self-consistency condition. 
This microscopic approach, while neglecting the
wave-vector dependences of the self-energies,
does encompass both the SDW and the local quantum critical
scenarios.} \label{fig1}
\end{figure}

An alternative to the traditional approach 
that is able to explain the aforementioned experiments 
is the local quantum critical scenario, in which 
a destruction of Kondo screening of the f-moments
coincides with the magnetic transition of the Kondo lattice~\cite{Si.01}.
The name local quantum criticality refers to the localization of the 
electronic excitations associated with the f-moments. It also stresses
the importance of long-range 
correlations along imaginary time, when the spatial correlations 
remain mean-field like as in the SDW scenario. 
Both types of QCP have been described with the extended 
dynamical mean field theory (EDMFT) 
approach to the Kondo lattice 
model, see Fig.\ref{fig1}.
 
\section{Extended DMFT, quantum critical Bose-Fermi Kondo models and the quantum-to-classical mapping}

The quantum critical point in the heavy fermion compounds arises out of the competition between
Kondo screening and the RKKY interaction~\cite{Doniach.77}. A minimal 
model for these systems
is the Kondo lattice model
\begin{equation}
  \label{eq:Kondolattice}
  H\,=\,\sum_{k,\sigma} \epsilon_k^{} d^{\dagger}_{k,\sigma}d^{}_{k,\sigma}
\,+\,
J_K \sum_{i} 
{\bf s}_{el}(\vec{r}_i) \cdot {\bf S}_i
\,+\,
\sum_{ ij} 
\frac{I_{ij}}{2} ~{\bf S}_{i} \cdot{\bf S}_{j}
~,
\end{equation}
where ${\bf s}_{el}(\vec{r}_i)$ denotes the conduction 
$d$-electron spin density at the location 
of the local moment ${\bf S}_i$.
The EDMFT approach
captures not only the Kondo screened Fermi liquid phase 
and the itinerant magnetic phase but also the dynamical
competition between the two underlying interactions.
It maps the Kondo lattice onto 
the Bose-Fermi Kondo (BFK) model, 
\begin{eqnarray}
{\cal H}_{\mbox{\small bfkm}} &=& J_K ~{\bf S}
\cdot {\bf s}_c + \sum_{p\sigma}
E_{p}~c_{p\sigma}^{\dagger}~ c_{p\sigma}
\nonumber\\
&+& \; 
g \sum_{p} {\bm S} \cdot \left( {\bm \phi}_{p} + {\bm \phi}_{-p}^{\;\dagger}
\right) + \sum_{p} w_{p}\,{\bm \phi}_{p}^{\;\dagger} {\bm \phi}_{p}\;,
  \label{eq:BFKM}
\end{eqnarray}
augmented by a self-consistency constraint, see Fig.\ref{fig1}.
 Here ${\bf S}$ is a spin-$1/2$ local moment, 
$J_K$ and $g$ 
are the Kondo coupling 
and the coupling constant to the bosonic bath respectively,
$c_{p\sigma}^{\dagger}$ describes a fermionic bath
with a constant density
of states, $\sum_{p} \delta (\omega - E_{p}) = N_0$,
and ${\bm \phi}_{p}^{\;\dagger}$ is the  bosonic bath
with the self-consistently determined spectral density.

Numerical evaluations
of the EDMFT scheme for the easy-axis Kondo lattice at finite
temperature~\cite{Grempel.03,Zhu.03} and at
zero temperature~\cite{Zhu.07,Glossop.07} indicate that the general
phase diagram of the Kondo lattice
contains both the Hertz-Millis and the local quantum critical point. 
Within EDMFT, the local quantum critical point of the lattice problem is related
to the quantum critical point (QCP) of a dissipative BFK model~\cite{Zhu.02,Zarand.02,Si.03}.
If the self-consistency condition leads to a
spectral density for the bosonic bath that behaves as
\begin{eqnarray}
\mbox{Im}\chi_0^{-1}(\omega)\,&\equiv&\,\sum_p [\delta(\omega-w_p)- \delta(\omega+w_p)]\nonumber \\ 
&\,\sim&\,
|\omega|^{1-\epsilon} \sgn(\omega) \Theta(\omega_c-|\omega|)
\label{EQ:sub-Ohmic}
\end{eqnarray}
below some dynamically generated scale $\omega_c$ with a bath exponent $0<\epsilon<1$, the
lattice QCP is mapped on the QCP of the sub-Ohmic BFKM. 

When arguments related to the ones that
lead to the
Ginzburg-Landau-Wilson functional of order-parameter fluctuations
are applied,
the  sub-Ohmic BFK model
would be mapped to a one-dimensional classical spin chain with long-ranged
interaction~\cite{Suzuki.76} and, by extension, 
a local-$\phi^4$ theory in $0+1$ dimension~\cite{Fisher.72,Suzuki.72}.
We will refer to this 
as the quantum-to-classical mapping of the QCP in the BFK model.
In general, 
the quantum-to-classical mapping relates the quantum criticality
to the classical criticality of order-parameter fluctuations
in elevated dimensions~\cite{Sachdev}.
For the local $\phi^4$ theory~\cite{Fisher.72,Suzuki.72}, 
$\epsilon=1/2$ effectively acts as an upper
critical dimension and
a Gaussian fixed point 
is expected for $1/2 < \epsilon < 1$. 
Correspondingly, there will be a violation of $\omega/T$
scaling due to the dangerously irrelevant quartic coupling
of the $\phi^4$ theory.
For $\epsilon=0$, the bosonic bath of the BFK model is ohmic and 
the corresponding classical spin chain is placed at its
lower critical dimension. The transition at $\epsilon=0$
is thus Kosterlitz-Thouless like~\cite{Anderson.69,Yuval.70}. 

This raises the question as to how EDMFT manages to yield critical properties
that differ from those of
a local $\phi^4$-theory, 
especially
the $\omega/T$-scaling of the order parameter susceptibility
~\cite{Grempel.03,Zhu.03}.
One possibility is that the self-consistency requirement leads effectively to
a different quantum impurity problem for every temperature, i.e. 
$\mbox{Im}\chi_0^{-1}=\mbox{Im}\chi_0^{-1}(\omega,T)$ is explicitly $T$-dependent
in contrast to Eq.~(\ref{EQ:sub-Ohmic}).
Another possibility is that already
the quantum critical point of the BFK model cannot be described by a local $\phi^4$-theory and that 
the quantum-to-classical mapping relating the two fails for the BFK 
model. This possibility has been recently discussed for the totally spin-isotropic BFK model as well as the
easy-axis BFK model~\cite{Zhu.04,Vojta.05,Glossop.05,Winter.08,Kirchner.08c,Kirchner.08d}.
As already mentioned, the QCP arises out of the competition between
the Kondo effect and magnetic fluctuations.
It therefore seems a prerequisite to choose an approach that correctly captures
Kondo screening, including the restoration of SU(2) invariance at the 
Kondo-screened fixed point.
For this reason the easy-axis BFK model
is 
delicate
~\cite{Winter.08,Kirchner.08c}.
In the following, we will focus on the  spin-isotropic
BFK model where the 
restoration of the SU(2) invariance in the Kondo-screened phase
is not an issue.

The starting point for the quantum-to-classical mapping is the
coherent state path integral representation of the quantum problem,
which leads to the most classical formulation of the
problem~\cite{Schulman}.
In the case of the BFK model, the effective
functional integral is formulated in 
terms of spin coherent states to take into account the finite size of the Hilbert space for the
local moment~\cite{Perelomov,Inomata}. 
\begin{figure}[t!]
\begin{center}
\includegraphics[width=0.5\textwidth]{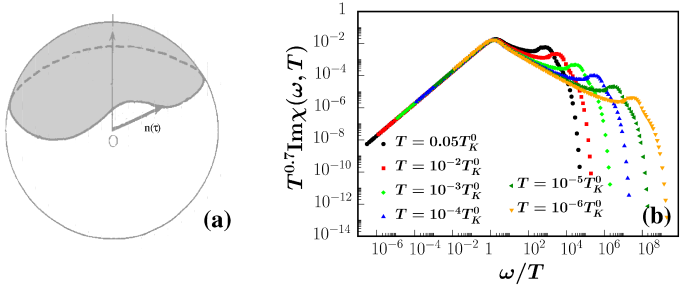}
\end{center}
\caption{(a) The parameter space for the spin path integral is the unit sphere S$^2$. The Berry phase
associated with a closed path is $s$ times the area traced out by $\vec{n}(\tau)$ 
($s$ is the size of the spin). For the path 
depicted in the figure the Berry phase is  $s$ times the shaded area.
(b) Scaling plot of the local spin susceptibility $\chi_{\mbox{\tiny loc}}(\omega,T)$
for $\epsilon=0.3$ at the critical coupling  $g_c/T_K^0=1.55$.
} \label{fig2}
\end{figure}
As a consequence, the order parameter manifold is the unit sphere 
in three dimensions and possesses
rotational invariance. The usual canonical term of the bosonic 
coherent state path integral
has to be  replaced by a topological term,
the Berry phase, see Fig.~\ref{fig2}. 
In Ref.~\cite{Kirchner.08d} it was  shown that 
the spin-isotropic BFK model can only be related 
to a local $\phi^4$-theory 
if the Berry phase
term of the spin coherent state path integral is neglected.  
In this case,
\begin{equation}
  \label{eq:dangirrel}
  \chi_{\mbox{\tiny loc}}(\omega,T=0)\sim \omega^{-(1-\epsilon)},~0<\epsilon<1.
\end{equation}
On the other hand,
due to the presence of a dangerously irrelevant
quartic coupling of the $\phi^4$ theory, the local spin 
susceptibility at criticality  for $\epsilon>1/2$ behaves as
\begin{equation}
  \label{eq:dangirrelT}
  \chi_{\mbox{\tiny loc}}(\omega=0,T)\sim T^{-1/2}.
\end{equation}
Comparing Eqs.~(\ref{eq:dangirrel},\ref{eq:dangirrelT})
shows the absence of the $\omega/T$-scaling.

The spin-isotropic BFK  model in the presence of the Berry phase term
has been studied in a dynamical large-$N$ limit~\cite{Zhu.04}. 
The Hamiltonian
of the SU(N)$\times$ SU($\kappa$ N) BFK model is
\begin{eqnarray}
{\cal H}_{\mbox{\tiny MBFK}} &=&
({J_K}/{N})
\sum_{\alpha}{\bf S}
\cdot {\bf s}_{\alpha}
+ \sum_{p,\alpha,\sigma} E_{p}~c_{p \alpha
\sigma}^{\dagger} c_{p \alpha \sigma}
\nonumber\\
&+&
({g}/{\sqrt{N}})
{\bf S} \cdot
{\bf \Phi}
+ \sum_{p}
w_{p}\,{\bf \Phi}_{p}^{\;\dagger}\cdot {\bf \Phi}_{p},
\label{H-MBFK}
\end{eqnarray}
where  $\sigma = 1, \ldots, N$ and
$\alpha=1, \ldots, \kappa N$ (with $\kappa N$ integer) 
are the spin and channel indices respectively,
and ${\bf \Phi} \equiv \sum_p ( {\bf \Phi}_{p} +
{\bf \Phi}_{-p}^{\;\dagger} )$ contains $N^2-1$
components.  
The local moment is  expressed in terms of pseudo-fermions
$S_{\sigma,\sigma^{\prime}}=f^{\dagger}_{\sigma}f^{}_{\sigma^{\prime}}
-\delta_{\sigma,\sigma^{\prime}}Q/N$,
where $Q$ is related to the chosen irreducible representation 
of SU(N)~\cite{Parcollet.98,Cox.93}.
At the QCP, the order parameter susceptibility~\cite{Zhu.04}
behaves as
\begin{equation}
\chi_{\mbox{\tiny loc}}(\omega,T=0)\sim 1/\omega^{1-\epsilon};~~ 
\chi_{\mbox{\tiny loc}}(\omega=0,T)\sim 1/T^{1-\epsilon}
\label{scal1}
\end{equation}
for all $0<\epsilon<1$.
The  $\omega/T$-scaling of $\chi_{\mbox{\tiny loc}}(\omega,T)$
for $1/2 \le \epsilon < 1$ implies the breakdown of the quantum
to classical mapping, since the mapped classical critical point
has $\chi_{\mbox{\tiny loc}}(\omega=0,T)\sim 1/T^{1/2}$ for  $1/2<\epsilon<1$.
That this large-$N$ result is stable against dangerously irrelevant couplings that may occur
at finite N has been demonstrated in~\cite{Kirchner.08d} for the particular value $\epsilon=2/3$.
This issue is addressed by introducing a
self-energy for the order parameter
susceptibility,
\begin{equation}
M(\omega,T)\equiv g_c^2\chi_0^{-1}(\omega) 
-1/\chi_{\mbox{\tiny loc}}(\omega,T),
\end{equation}
where $\chi_0^{-1}(\omega)$ follows from Eq.~(\ref{EQ:sub-Ohmic}) 
and $g^2_c(J_K^{})$ is the value of the
coupling constant $g^2$ at which the system becomes 
critical for a given $J_K^{}$. 
The $\omega$-independent part of $M(\omega,T=0)$, $M(\omega=0,T=0)$,
  vanishes at  $g^2_c(J_K^{})$ by definition.

\begin{figure}[t!]
\begin{center}
\includegraphics[width=0.5\textwidth]{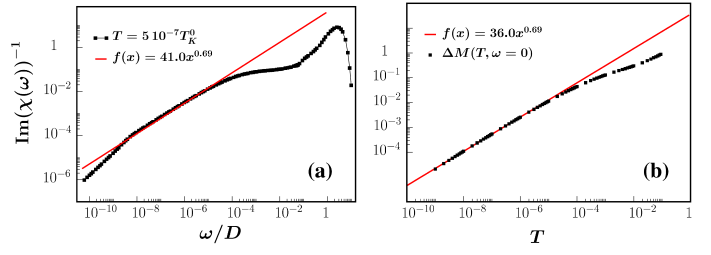}
\end{center}
\caption{(a) The imaginary part of the inverse of the order parameter susceptibility. It displays
power-law behavior as a function of $\omega$ with the same exponent as $\chi^{-1}_{0}(\omega)$ over
a region that is cut off by the temperature of the system ($T_K^0=0.05 D$). 
(b) The temperature dependent part of the saddle point self-energy displays power-law
behavior over roughly five decades: $\Delta M(\omega=0,T)\sim T^{\alpha}$, where $\alpha\approx 1-\epsilon$ and
$\epsilon=0.3$. } 
\label{fig3}
\end{figure}
This self-energy has a temperature dependent part, formally defined as
\begin{equation}
\Delta M(\omega,T)=M(\omega,T)-M(\omega=0,T=0),
\end{equation}
with the important property
\begin{equation}
\Delta M(\omega=0,T)\sim T^{1-\epsilon}.
\end{equation}
It was shown in Ref.~\cite{Kirchner.08d} that 
this temperature dependence implies that no dangerously irrelevant coupling can alter
$\chi_{\mbox{\tiny loc}}(\omega=0,T)\sim 1/T^{1-\epsilon}$
 for  $\epsilon>1/2$.

We will now demonstrate that a similar spin self-energy can be obtained for the region
with  $\epsilon<1/2$. Numerical details can be found in Ref. \cite{Zhu.04,Kirchner.08d}.
The dynamical spin susceptibility obeys the scaling form
\begin{equation}
\chi_{\mbox{\tiny loc}}(\omega,T)=T^{\epsilon-1}\Phi(\omega/T),
\label{scal2}
\end{equation}
as demonstrated in Fig.\ref{fig2}b 
for the particular value $\epsilon=0.3$ and consistent with Eq.~(\ref{scal1}).
The  imaginary part of the inverse of the spin susceptibility is therefore $\omega$-dependent with
a power-law consistent with that for $\mbox{Im}\chi_0^{-1}(\omega)$ from Eq.(\ref{EQ:sub-Ohmic}).
This is shown in Fig.\ref{fig3}a for a particular temperature ($T=5 \times 10^{-7} T_K^0=2.5 \times 10^{-8}D$,
where $D$ is the half-bandwidth). The resulting spin self-energy at the critical coupling $g_c$
is shown in  Fig.\ref{fig3}b. It behaves as
$\Delta M(\omega=0,T)\sim T^{\alpha},~\alpha=0.69\approx 1-\epsilon$. 
This demonstrates again that the critical properties of the spin-isotropic
BFK model (i.e. with the Berry phase) 
have the same characteristics 
for both $0<\epsilon<1/2$ and $1/2<\epsilon<1$.
Finally, we note in passing that the critical point of the 
$\epsilon=1/2$ case does not appear to be special compared to
that of the $0<\epsilon<1/2$ and $1/2<\epsilon<1$ cases.
In particular,
we do not observe the emergence at $\epsilon=1/2$
of any logarithmic corrections to the 
power laws
in the dynamical local susceptibility and related quantities.

\section{Summary}
In summary, the current interest in quantum critical heavy fermion 
systems has led to the question as to whether the Kondo effect itself
may become critical concomitant with the magnetic quantum transition.
The effect of a critical Kondo destruction can already be studied in 
a  sub-Ohmic
Bose-Fermi Kondo impurity model. We have reexamined the evidence that
the Berry phase term cannot be neglected at the quantum critical 
point and that
its presence changes the critical properties. Without the Berry phase term 
the quantum criticality 
is in the universality class
of the classical local $\phi^4$-theory.  
With the Berry phase term 
the quantum critical point is interacting in the whole
parameter region $0<\epsilon<1$ and,
therefore, not equivalent to that of a local $\phi^4$-theory. 
We demonstrated that the temperature dependence of the
spin self-energy has the same characteristics 
between $0<\epsilon<1/2$ and $1/2<\epsilon<1$.
Since the standard quantum-to-classical 
mapping links
the critical properties of sub-Ohmic impurity models to 
those of one-dimensional classical 
spin-chains
with long-ranged interaction,
one is led to the conclusion that this mapping fails for
the sub-Ohmic Bose-Fermi Kondo model.

\section{Acknowledgments}
We 
thank C.~J.~Bolech for useful discussions.
The work has been supported  in part by the
NSF I2CAM International 
Materials Institute Award, Grant No. DMR-0645461, the
NSF Grant No. DMR-0706625, the Robert A. Welch Foundation,
the W. M. Keck Foundation,
the Rice Computational Research Cluster
funded by NSF
and a partnership between Rice University, AMD and Cray,
and (for S.K.) DOE Grant No. DE-FG-02-06ER46308.


\end{document}